\documentclass{llncs}
\usepackage{mathptmx}       
\usepackage{helvet}         
\usepackage{courier}        
\usepackage{type1cm}        
\usepackage{makeidx}         
\usepackage{graphicx}        
\usepackage{multicol}        
\usepackage[bottom]{footmisc}
\usepackage{subfigure}
\usepackage{amsfonts}
\usepackage[cmex10]{amsmath}
\usepackage{hyperref}
\usepackage{amssymb}
\usepackage{algorithm}
\usepackage{algpseudocode}

\begin{document}

\title{Beyond One Solution: The Case for a Comprehensive Exploration of Solution Space in Community Detection}
\titlerunning{Exploration of Solution Space ...}  
%
\author{Fabio Morea\inst{1} \and Domenico De Stefano \inst{2}}
\authorrunning{Morea, F. et al.} 

\institute{Area Science Park, Padriciano 99, Trieste, Italy \email{fabio.morea@areasciencepark.it}
\and
University of Trieste,
Piazzale Europa 1, Trieste, Italy \email{ddestefano@units.it}}

\maketitle    
\section{Introduction and methodology}
This article explores the importance of examining the solution space in community detection, highlighting its role in achieving reliable results when dealing with real-world problems. A methodology and a taxonomy are proposed to describe different types of solution spaces. 

\begin{figure} [h]
    \centering
    \includegraphics[width=1.0\linewidth]{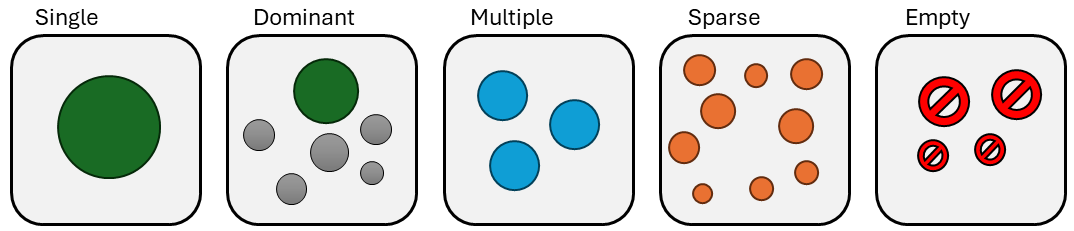}
     \caption{Taxonomy for the solution space of generated by a community detection algorithm.}
     \label{fig:taxo}
\end{figure}

Let $G=(V,E)$ be a graph, with $n_v = |V|$ vertices and $n_e = |E|$ edges. A \textbf{community} $C$ is defined as a subnetwork of $G$ that satisfies a condition: nodes that belong to $C$ are more densely connected within each other than with the rest of $G$. A \textbf{partition} P represents a set of $k$ disjoint subnetworks $C_1, \ldots, C_k$ whose union is equal to $G$.
A community detection algorithm $\mathcal{A}(G, \rho) \rightarrow P$ is a function that takes as input a graph $G$ and one or more parameters $\rho$, and returns a partition $P$. Several community detection algorithms are discussed in literature \cite{Diboune2024,Khawaja2024}. Ideally, any of them should produce a single, valid partition each time it is applied with the same parameters. In practice, however, for large, dense networks, $\mathcal{A}$ may produce different partitions, $P_i \ne P_j$ at each trial, or may generate invalid partitions. 
The \textbf{solution space} $\mathbb{S} = \{ P_1, P_2, \ldots, P_{ns} \}$ is the set of all unique partitions that $\mathcal{A}$ produces across $t$ trials. Our research aims to determine the minimum number of trials $t_c$ required to confidently assert that $\mathbb{S}$ is \textit{stable}, i.e. it is unlikely to expand with an additional run of $\mathcal{A}$. Moreover, we introduce a taxonomy to classify $\mathbb{S}$ in different categories (depicted in Figure \ref{fig:taxo}), based on $ns$ and the relative frequencies of $P_i$ observed in the trials. 

To understand whether $\mathbb{S}$ is stable and the relative importance of each of the solutions found, we created an experimental setting, described in Algorithm 1, to produce a probability model $\mathbb{M}$ under a Bayesian framework. The occurrence of new solutions is modeled as a series of Bernoulli trials, where each trial results in either a "success" (a new solution is found) or a "failure" (an identical solution is observed). $\mathbb{M}$ is a Beta-Binomial model, initialised with a non-informative prior. As trials progress, $\mathbb{M}$ is updated. The probability that   $\mathbb{S}$ is stable after $t$ trials (meaning the $t+1$ trial will not yield a new solution) can be modeled as $p_{stable} = 1-\mathbb{E}(\beta(t+2, t-ns+2)) $, where $ns$ is the number of solutions found after $t$ trials and $\mathbb{E}$ is the mean of $\beta$ distribution. The process continues until either $t_{max}$ is reached or the $p_{stable}$ reaches a predefined threshold $\tau$.
From the model $\mathbb{M}$ we can derive point estimates $\bar{p}_i$ and interval estimates $p_{i,lower}, p_{i,upper}$ of the frequency if any solution $P_i$.   

\begin{algorithm}
\caption{Solution Space Exploration}
\begin{algorithmic}[1]
\State \textbf{Input:} Graph $G$, algorithm $\mathcal{A}$, $t_{max}$, $\tau$
\State \textbf{Initialize:} Empty solution space $\mathbb{S} \leftarrow \emptyset$
\State \textbf{Initialize:} Non-informative prior for Beta-Binomial model $\mathbb{M} \leftarrow \text{Beta}(1,1)$
\For{$i = 1$ to $t_{max}$}
    \State Shuffle the network $G$ to generate a random permutation $G^*$
    \State Run community detection algorithm $\mathcal{A}(G^*, \rho)$ to obtain partition $P_i$
    \If{$P_i \notin \mathbb{S}$}  
        \State Add the new solution $P_i$ to solution space $\mathbb{S}$ 
        \State Add the new solution $P_i$ to Bayesian model $\mathbb{M}$  
    \EndIf
    \State Bayesian update of $\mathbb{M}$
    \State \textbf{If} probability of $P_{stable} > \tau$ \textbf{then} exit the loop  
\EndFor
\State \textbf{Output:} Solution space $\mathbb{S}$ and Bayesian model $\mathbb{M}$
\end{algorithmic}
\end{algorithm}

An important step in the exploration of $\mathbb{S}$ is highlighted in step 4 of the algorithm: $G$ should be permuted at each trial, to avoid incurring in the \textit{input ordering bias}, as discussed in \cite{morea2024CCD}. 
Moreover, a partition $P$ may be considered \textbf{invalid} for several reasons: it may be trivial (e.g., $k = 1$ or $k = n_v$), internally disconnected, or fail to meet the community definition (i.e. nodes in $C_i$ are more connected to nodes in any other partitions $C_j$ (where $j \neq i$) than within $C_i$.  
The taxonomy proposed in Figure \ref{fig:taxo} can be formally defined as follows: the \textit{Single} category describes the case where the solution space is stable and there is only one valid partition ($ns = 1$). The \textit{Dominant} category occurs when there are multiple valid partitions ($ns > 1$), but one partition is dominant, i.e., $\max(p_{lower}) > 0.5$. The \textit{Multiple} category applies when $ns > 1 $ and $max(p_{lower}) < 0.5$. The \textit{Sparse} category occurs when a high number of solutions exists ($n_s \approx t$) each with a low probability ($max(p_{upper}) \approx 0$). Lastly, the \textit{Empty} category represents a situation where there are no solutions, or all solutions are invalid, i.e $ns = 0$.

\section{Results and conclusions}
\begin{figure}
     \centering
     \includegraphics[width=0.99\linewidth]{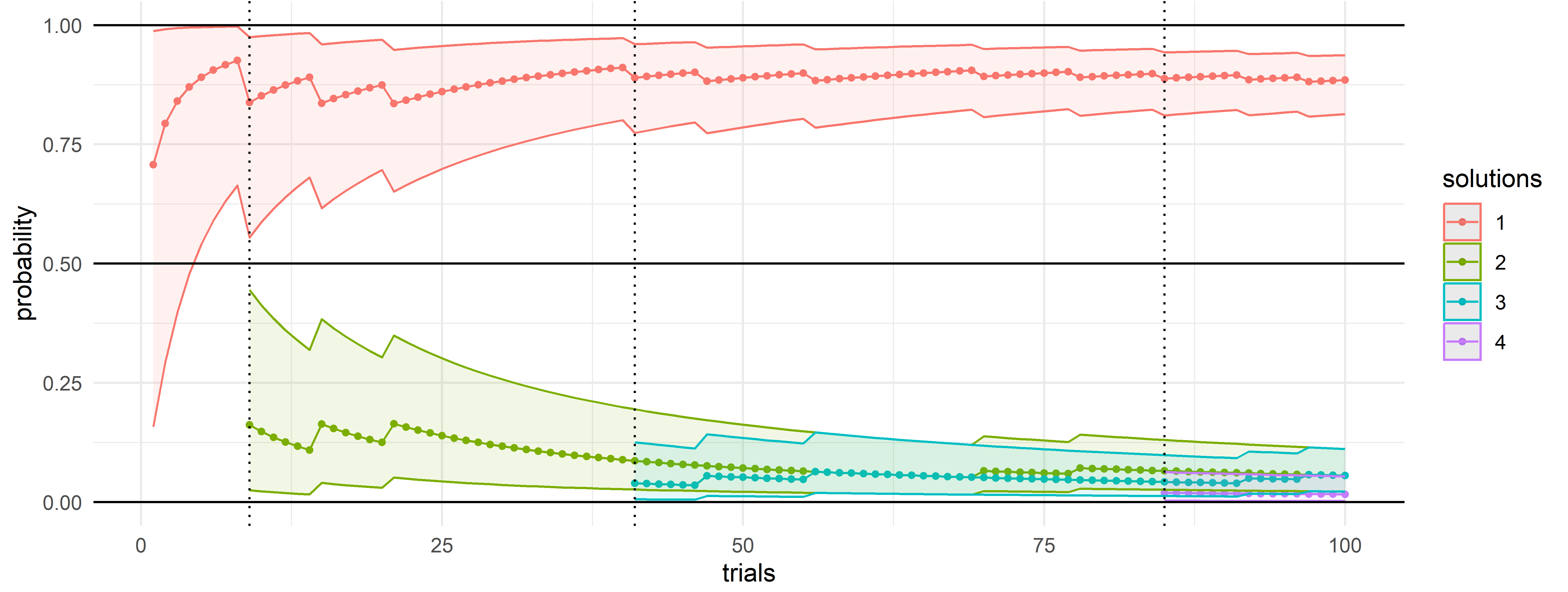}
     \caption{Confidence intervals associated with different solutions in the solution space.}
     \label{fig:confidence}
 \end{figure}

To illustrate the methodology, examples are presented by applying it to a large, dense network from the "Horizon Projects Network" dataset, which describes collaborations between organizations in EU-funded research projects from 2015 to 2029. The network $G_{2024}$ is partitioned using Infomap algorithm \cite{Rosvall_2008}. The evolution of knowledge regarding the solution space as a function of the number of trials is illustrated in Figure \ref{fig:confidence}: the probability of solutions in $\mathbb{S}$ are represented as point estimates (solid line for $\bar{p}$) and intervals (ribbon between $(p_{lower}$ and $p_{upper}$). A dominant solution (shown in red) emerges soon; as $t$ increases additional solutions are discovered and beyond $t=50$, the probability distributions are unaffected by any new solutions. When the experiment is repeated, a slightly different situation may observed, but convergence towards a dominant solution consistently occurs.

\begin{figure}
     \centering
     \includegraphics[width=0.99\linewidth]{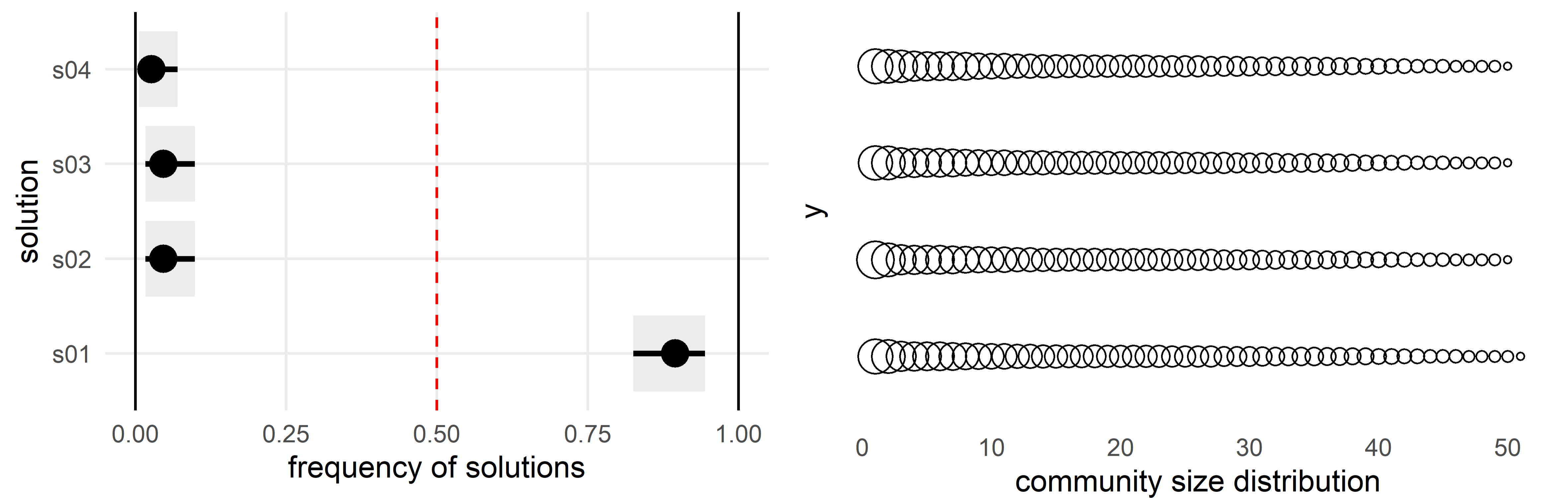}
     \caption{Example of a Dominant solution space: $ns = 4$, $max(p_{lower}) > 0.5$}
     \label{fig:sspIM}
 \end{figure}

The corresponding solution space $\mathbb{S}^{IM}$ falls into the dominant solution with $ns = 4$ and $max(p_i) \approx 0.85$ and its main features are shown in Figure \ref{fig:sspIM}. The community size distribution for each partition is depicted in the diagram on the right, highlighting that all solutions have $k = 55$. 

\begin{figure}
     \centering
     \includegraphics[width=0.99\linewidth]{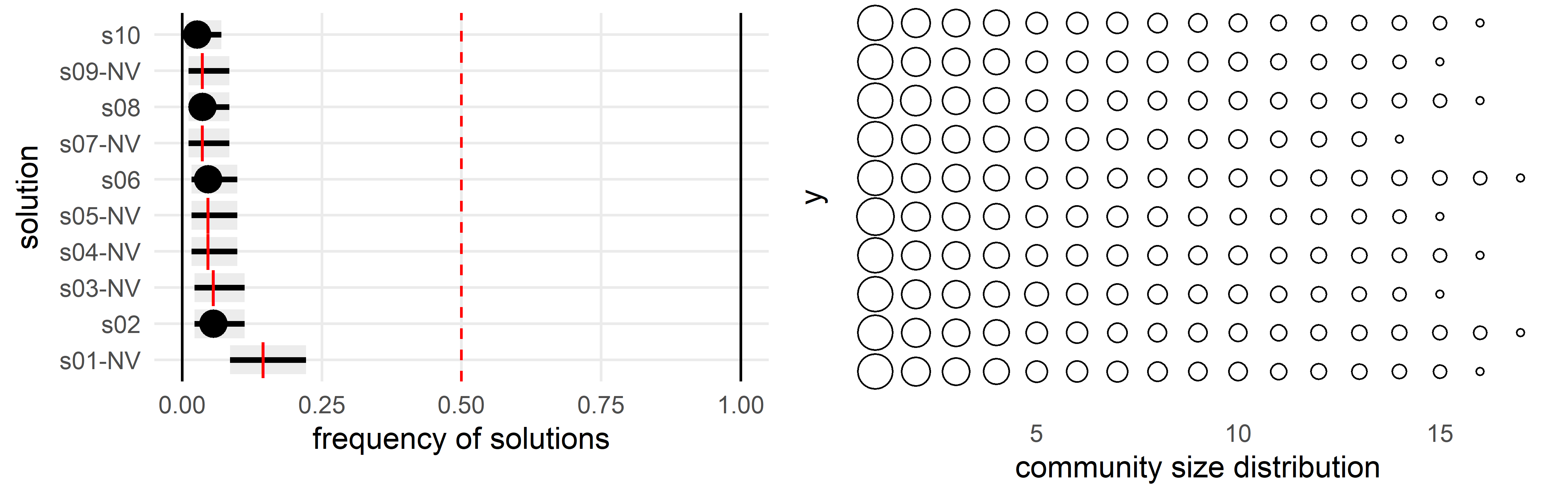}
     \caption{Example of a Sparse solution space: $ns = 10$ and $max(p_{upper}) < 0.5$}
     \label{fig:sspLV}
 \end{figure}

Partitioning the same network with Louvain \cite{Blondel_2008} algorithm generates different solution space $\mathbb{S}^{LV}$ with $ns = 46$, and $max(p_i) \leq 0.25$ that falls in the Sparse category of our taxonomy. In this case $k$ varies across solutions, and most of the solutions are invalid due to the presence of disconnected communities. Figure \ref{fig:sspLV} shows the main characteristics $\mathbb{S}^{LV}$; for the purpose of clarity the figure is limited to the top 10 solutions). The methodology and examples above highlights the importance of exploring the solution space, rather than relying on a single detected partition, as multiple valid solutions may exist.

If the solution space falls in the Single or Dominant category, it can be analysed in a straightforward way. However, in the case of multiple solutions it may be necessary to take focus on those with the highest probability and proceed with a consensus approach such as in \cite{morea2024CCD}. If the solution space is Sparse, the number of solutions to be taken into account is higher, and even consensus may result in unreliable results: it is the case it may be advisable to examine similarity between solutions, or to simplify the network before community detection. Utilizing a Bayesian framework to assess the structure of $\mathbb{S}$ is useful to determine when further exploration is unlikely to yield new solutions: this ensures a thorough yet efficient analysis, and optimizes computational resources.

---

\textbf{Data and code availability:} all data and code utilized in this study are available under a CC-BY license. The sample network $G_{2024}$ is part of the Horizon Projects Network dataset (DOI \href{https://zenodo.org/records/13594210}{10.5281/zenodo.13594209}). The analysis was conducted using the R programming language and packages \textit{igraph} (\href{https://r.igraph.org/}{https://r.igraph.org/}) and \textit{communities} (\href{https://github.com/fabio-morea/communities}{https://github.com/fabio-morea/communities}).

\bibliographystyle{splncs03} 
\bibliography{refs} 

\begin{thebibliography}{1}
\providecommand{\url}[1]{\texttt{#1}}
\providecommand{\urlprefix}{URL }

\bibitem{Blondel_2008}
Blondel, V.D., Guillaume, J.L., Lambiotte, R., Lefebvre, E.: Fast unfolding of communities in large networks. Journal of Statistical Mechanics: Theory and Experiment  2008(10),  P10008 (Oct 2008), \url{http://dx.doi.org/10.1088/1742-5468/2008/10/P10008}

\bibitem{Diboune2024}
Diboune, A., Slimani, H., Nacer, H., Bey, K.B.: A comprehensive survey on community detection methods and applications in complex information networks. Social Network Analysis and Mining  14(1), ~93 (2024), \url{https://doi.org/10.1007/s13278-024-01246-5}

\bibitem{Khawaja2024}
Khawaja, F.R., Zhang, Z., Memon, Y., Ullah, A.: Exploring community detection methods and their diverse applications in complex networks: a comprehensive review. Social Network Analysis and Mining  14(1),  115 (June 2024), \url{https://doi.org/10.1007/s13278-024-01274-1}

\bibitem{morea2024CCD}
Morea, F., DeDtefano, D.: Enhancing stability and assessing uncertainty in community detection through a consensus-based approach. ArXiv prepring  (2024), \url{https://doi.org/10.48550/arXiv.2408.02959}

\bibitem{Rosvall_2008}
Rosvall, M., Bergstrom, C.T.: Maps of random walks on complex networks reveal community structure. Proceedings of the National Academy of Sciences  105(4),  1118–1123 (Jan 2008), \url{http://dx.doi.org/10.1073/pnas.0706851105}

\end{thebibliography}

\end{document}